\documentstyle[aps,prl,epsfig]{revtex}
\begin{document}
\twocolumn[\hsize\textwidth\columnwidth\hsize
          \csname @twocolumnfalse\endcsname

\title{Asymptotic limit of high spatial dimensions and thermodynamic
  consistence} 

\author{V.  Jani\v{s}}

\address{Institute of Physics, Academy of Sciences of the Czech Republic,\\
  Na Slovance 2, CZ-18221 Praha 8, Czech Republic}

\date{\today}
\maketitle 
\begin{abstract}
  The question of thermodynamic consistence and $\Phi$-derivability of the
  asymptotic limit of high spatial dimensions for quantum itinerant models
  is addressed. It is shown that although the irreducible $n$-particle
  Green functions are local, reducible vertex functions retain different
  momentum dependence. As a consequence, the vertex corrections to
  conductivity do not generally vanish in the mean-field limit.  The
  mean-field theory is a $\Phi$-derivable approximation only if regular
  nonlocal or anomalous local external sources are admitted.\\
  PACS numbers: 71.10.-w, 71.10.Fd\\
\end{abstract}

] 

The Baym and Kadanoff concept of thermodynamic consistence and conserving
character of approximations is derived from an observation that knowing a
thermodynamic potential as a functional of external disturbances we can
reconstruct thermally averaged values of any observable \cite{Baym61}. The
most transparent way to guarantee thermodynamic consistence is to find a
generating Luttinger-Ward functional $\Phi\left[G_{J}\right]$ depending on
the fully renormalized one-particle propagator $G_{J}$ in the presence of a
set of external disturbances $J_\alpha$. Functional derivatives of the
generating functional with respect to $G$ lead to irreducible functions,
while functional derivatives with respect to the external sources
$J_\alpha$ to reducible ones \cite{Baym62}. The $n$-th orders of the
derivatives generate $n$-particle functions, either irreducible or
reducible.  This definition of higher-order Green functions guarantees a
conserving character of the approximation. But it may be in conflict with
criteria for selecting Feynman diagrams contributing to the $n$-particle
Green functions in special limits such as the mean-field theory. It is
defined as an exact solution in infinite spatial dimensions.

Selection of relevant diagrams in high spatial dimensions is determined by
a scaling of nonlocal terms in lattice Hamiltonians.  It can, from
principal reasons, be made only within a perturbation expansion. In lattice
models of correlated and/or disordered electrons one can conclude from
finiteness of the energy density on a collapse of diagrams to a single
site, \cite{Metzner89} or on irrelevance of momentum conservation at
internal vertices \cite{Mueller-Hartmann89}.  The one-electron self-energy
gets local in infinite spatial dimensions and can be generated from a
Luttinger-Ward functional depending on only local elements of the full
one-particle propagator. Such a functional, or its Legendre transform was
constructed for Hubbard-like models without and with chemical randomness in
\cite{Janis91,Janis92b}.  As far as we are interested only in the dynamics
of one-particle quantities, we can completely neglect the off-diagonal
elements of the propagator \cite{Georges96}.

The limit of high spatial dimensions has an asymptotic character, i.~e.
leading orders in the small parameter $d^{-1}$ are to be kept for each
quantity. The off-diagonal elements of the one-particle propagator scale on
a hypercubic lattice as $d^{-|{\bf i}-{\bf j}|/2}$, where $|{\bf i}-{\bf
  j}|$ the lattice distance between sites ${\bf i}$ and ${\bf j}$. The
off-diagonal elements are essential for susceptibilities deciding about the
existence of solutions with a long-range order \cite{Brandt89}.

To prove a long-range order we have to evaluate two-particle
susceptibilities (correlation functions) at fixed momenta. The collapse of
diagrams in the mean-field limit is well defined in the lattice space but
the corresponding reduction in the momentum dependence of two-particle
Green functions, depending generally on three momenta, is less evident.
Particularly, if the limit $d\to\infty$ is used within the Baym and
Kadanoff conserving scheme, the locality in two-particle functions is still
understood controversially \cite{Engelbrecht95,vanDongen98}.

It is the aim of this Letter to address the problem of compatibility of the
high-dimensional asymptotics with the Baym concept of $\Phi$-derivability
of all Green functions.  We show that all \textit{ nonperturbative
  irreducible} vertex functions are local and can be generated from the
Luttinger-Ward functional via derivatives w.r.t. local propagators.
Reducible $n$-particle Green functions are generally nonlocal, but the
nonlocality is perturbative. They can be derived from the generating
functional if we either keep the external sources nonlocal or allow for
local anomalous perturbations that do not conserve charge or spin.

We can construct the limit to infinite lattice dimensions within the
Baym and Kadanoff scheme. We start with a general model the partition
function of which can be represented via a functional integral    
\begin{eqnarray} \label{eq:Z-int}
  \lefteqn{{\sf Z}\left\{J; G^{(0)-1}\right\} =}\\
  &&\int {\cal D}\varphi {\cal D}
 \varphi ^{*}\exp \left\{ -\varphi ^{*}\eta G^{(0)-1}\varphi +\varphi 
  ^{*}J\varphi +U\left[ \varphi ^{*} ,\varphi \right] \right\} \nonumber
\end{eqnarray}
where, for the sake of simplicity, we suppressed all decoration indices for
the lattice and internal degrees of freedom. All the interaction is
contained in a \textit{local} functional $U$, $J$ stands for external
sources, and $\eta=\pm 1$ for bosons, fermions, respectively.

A renormalized perturbation theory is achieved if we replace the bare
one-particle propagator $G^{(0)}$ with ${\cal G}=(G^{-1}+\Sigma)^{-1}$. It
is a straightforward task to write down a generating functional for
the renormalized perturbation theory. The grand potential contains the
renormalized one-particle propagator $G$ and the self-energy $\Sigma$ as
variational functions as follows
\begin{eqnarray}
  \label{eq:gen-fun}
  \Omega\left[J;G,\Sigma\right]&=&
  -\beta^{-1}\ln{\sf Z}\left\{0;G^{-1}+\Sigma\right\} +\eta\beta^{-1}
  \mbox{tr}\ln G \nonumber\\
  && +\eta\beta^{-1}\mbox{tr}\ln\left[G^{(0)-1}-\eta J-\Sigma\right]\, .  
\end{eqnarray}
The physical values of $G$ and $\Sigma$ are obtained from the equations
$\delta\Phi/\delta\Sigma=0$ and $\delta\Phi/\delta G=0$, respectively.
Note that $G$ and $\Sigma$ are Legendre conjugate variables.

Representation (\ref{eq:Z-int}) holds in any dimension. In the limit
$d\to\infty$ we have to separate site-diagonal and off-diagonal parts of
$G$ and $\Sigma$.  The electron systems show the following asymptotics
in high dimensions
\begin{equation}
  \label{eq:vj-diag-off}
  G=G^{diag}[d^0]+G^{off}[d^{-1/2}]\, ,\,\, \Sigma=\Sigma^{diag}[d^0]
  +\Sigma^{off}[d^{-3/2}] \, .
\end{equation}
The grand potential in the limit $d=\infty$ is obtained if the
off-diagonal parts $G^{off}$ and $\Sigma^{off}$ are completely neglected in
(\ref{eq:gen-fun}). The nonlocal part of the grand potential is
contained solely in the bare propagator $G^{(0)}$. Hence, all nonlocal
quantities in the mean-field theory are treated perturbatively without
renormalizations.

According to Baym and Kadanoff, $n$-particle Green functions can
be generated via successive applications of functional derivatives
w.r.t. the external source $J$. Local sources are  dominant in $d=\infty$
and we have for one- and  two-particle functions
\begin{equation}
  \label{eq:GFJ}
  G_{\bar{1}1'}[J]=-\beta\frac{\delta\Omega[J]}{\delta J_{\bar{1}1'}}=\langle 
  \varphi^*_{\bar{1}}\varphi^{\phantom*}_{1'}\rangle_J \, ,\,\, 
   L_{\bar{1}1',\bar{2}2'}=\eta\frac{\delta G_{\bar{1}1'}}{\delta
     J_{\bar{2}2'}}
\end{equation}
where $1\equiv ({\bf R}_1,t_1,\sigma_1,\ldots)$ denotes a set of degrees of
freedom in the space-time representation. The bar indicates a coordinate of
a complex-conjugate field and any decoration of the labeling index refers
always to the undecorated lattice site.

We now find equations of motion for two-particle functions and an explicit
representation for the two-particle vertex in $d=\infty$.  As a first step
we subtract the free two-particle propagation from
$L_{\bar{1}1',\bar{2}2'}$ and define a new vertex function
\begin{eqnarray}
  \label{eq:vertex}
  \Gamma_{\bar{1}1',\bar{2}2'}&=&\sum_{1'',1''',2'',2'''}G^{-1}_
  {\bar{1}1''}G^{-1}_{\bar{1}'''1'}\left[L_{\bar{1}''1''',\bar{2}''2'''}
   \right. \nonumber\\ && \left.-\eta G_{\bar{1}''2'''}G_{\bar{2}''1'''}\right]
  G^{-1}_{\bar{2}2''}G^{-1}_{\bar{2}'''2'}\, 
\end{eqnarray}
where the inversions use only the site-diagonal functions.

Analogously we define a local vertex $\gamma_{\bar{1}1',\bar{1}"1'''}$ 
from an averaged cumulant two-particle function 
\begin{eqnarray}
  \label{eq:gam-def}
 && \gamma_{\bar{1}1',\bar{1}''1'''}=\eta\!\!\! \sum_{2,2',2'',2'''}\!\!\!
 \delta_{{\bf R}_1,{\bf R}_2}G^{-1}_{\bar{1}2}G^{-1}_{\bar{2}'1'}
 \left\{\langle\varphi^*_{\bar{2}}\varphi^{\phantom*}_{2'}\varphi^*_{
     \bar{2}''}\varphi^{\phantom*}_{2'''}\rangle \right.\nonumber\\
  &&\left. -\langle\varphi^*_{\bar{2}}\varphi^{\phantom*}_{2'}\rangle
  \langle\varphi^*_{\bar{2}''}\varphi^{\phantom*}_{2'''}\rangle
  -\eta\langle\varphi^*_{\bar{2}}\varphi^{\phantom*}_{2'''}\rangle \langle 
  \varphi^*_{\bar{2}''}\varphi^{\phantom*}_{2'}\rangle\right\}
  G^{-1}_{\bar{1}''2''}G^{-1}_{\bar{2}'''1'''}\, . 
\end{eqnarray}
We use the above definitions in the mean-field limit of the generating
functional (\ref{eq:gen-fun}) with inhomogeneous but local one-particle
functions to find an explicit form of a two-particle function
$\delta\Sigma_{\bar{1}1'}/\delta J_{\bar{2}2'}$ needed for the evaluation
of $L_{\bar{1}1',\bar{2}2'}$. If we switch from $L$ to $\Gamma$ we finally
obtain a Bethe-Salpeter integral equation of motion
\begin{eqnarray}
  \label{eq:vertex-BS}
  \sum_{3,3'}&&\left\{\delta_{3,1}\delta_{3',1'}-\eta\sum_{1'',1'''}
    \gamma_{\bar{1}1',\bar{1}''1'''}G^{off}_{\bar{1}'''3}G^{off}_{\bar{3}'
      1''}\right\}\Gamma_{\bar{3}3',\bar{2}2' } \nonumber \\
   & &=\delta_{{\bf R}_1,{\bf R}_2}\gamma_{\bar{1}1',\bar{2}2'}     \, .
\end{eqnarray}
This form of the Bethe-Salpeter equation is natural for the mean-field
limit, since it separates diagonal and off-diagonal propagators. The former
are contained nonperturbatively in the local vertex $\gamma$ while the
latter participate only via two-particle bubbles forming the nonlocal part
of the vertex $\Gamma$.

We can further introduce a two-particle irreducible vertex that we denote
$\Lambda$. This function enables us to replace in (\ref{eq:vertex-BS}) the
off-diagonal one-particle propagators with the unrestricted ones. A new
equation of motion with the two-particle irreducible vertex reads
\begin{eqnarray}
  \label{eq:BS-lambda}
   \sum_{3,3'}&&\left\{\delta_{3,1}\delta_{3',1'}-\eta\sum_{1'',1'''}
    \Lambda_{\bar{1}1',\bar{1}''1'''}G_{\bar{1}'''3}G_{\bar{3}'1''}\right\}
  \Gamma_{\bar{3}3',\bar{2}2' }\nonumber\\
  &&=\delta_{{\bf R}_1,{\bf R}_2}\Lambda_{\bar{1}1',\bar{2}2'}\ .   
\end{eqnarray}
This form of the Bethe-Salpeter equation is common in general treatments
without the collapse of diagrams.  In the mean-field limit, $d=\infty$, the
irreducible vertex $\Lambda$ is the same for $\gamma$ as well as for
$\Gamma$.

The irreducible vertex function is related to the self-energy via a
generalized Ward identity
\begin{eqnarray}
  \label{eq:IR-var}
  \Lambda_{\bar{1}1',\bar{2}2'}&=&\delta_{{\bf R}_1,{\bf R}_2}
  \frac{\delta\Sigma_{\bar{1}1'}}{\delta G_{\bar{2}2'}}
\end{eqnarray}
following directly from the above derivation. Since the irreducible vertex
functions appear in integral kernels of the Bethe-Salpeter equations of
motion for higher-order Green functions, they get local in the lattice
space in the mean-field limit and hence independent of momenta in
spatially homogeneous solutions. Variations at fixed momenta as used in
\cite{Engelbrecht95,vanDongen98} loose their meaning where off-diagonal
elements in the lattice space are treated perturbatively. Nonlocal
corrections to the irreducible vertex functions are negligible in leading
order of $d\to\infty$.

The above derivation of the two-particle vertex function within the Baym and
Kadanoff conserving scheme is a generalization of the Brandt and Mielsch
construction of charge susceptibilities in the Falicov-Kimball model
\cite{Brandt89}. The review on the $d=\infty$ approach \cite{Georges96}
uses, however, a different construction of two-particle functions. The
question is whether the two constructions lead to the same vertex function.
In other words, does $\Gamma_{\bar{1}1',\bar{2}2' }$ contain all leading
$d^{-1}$ contributions to the two-particle vertex?

\begin{figure}
  \epsfig{figure=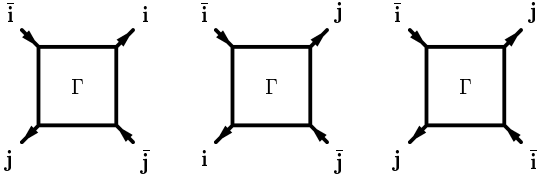,width=75mm}
\caption{\label{fig:2P-vertices} Leading order of two-particle
  vertex functions in $d\to\infty$. The choices of two distant lattice
  sites $\mathbf{i}, \mathbf{j}$ correspond to different two-particle
  reducibility channels, interaction ($U$), electron-hole ($eh$), and
  electron-electron ($ee$), respectively.}
\end{figure}

It is undisputable that maximally two different lattice sites are relevant
for two-particle functions in $d=\infty$ \cite{note0}. Two-particle
Green functions have four end points. We generally have three possibilities 
how to select couples of end points from the same lattice site, 
Fig.~\ref{fig:2P-vertices}. 
Attaching end points of two-particle functions to lattice sites
determines multiple repetition of nonlocal two-particle bubbles in the
Bethe-Salpeter equations for two-particle reducible functions. Each choice
in Fig.~\ref{fig:2P-vertices} is of the same order in the parameter
$d^{-1}$ and represents a different two-particle reducibility.  The first
choice, resulting from the above Baym and Kadanoff construction, leads to
the so-called interaction two-particle channel of a chain of spin-triplet
polarization bubbles screening the bare interaction. The second type of
nonlocality in the two-particle function induces electron-hole, and the
last one then electron-electron (spin-singlet) multiple scatterings.

We use the three ways of constructing nonlocal two-particle functions and
define three types of multiplication of the off-diagonal bubbles and the
local vertex in the Bethe-Salpeter equation (\ref{eq:vertex-BS})
\begin{mathletters}\label{eq:multiplication}
\begin{eqnarray}
  \label{eq:U-mult}
  \kappa^U_{\bar{1}1',\bar{2}2'}&=& \sum_{1'',1'''}\gamma_{\bar{1}1',
    \bar{1}''1'''}G^{off}_{\bar{1}'''2'} G^{off}_{\bar{2}1''}\, ,\\
  \label{eq:eh-mult} 
 \kappa^{eh}_{\bar{1}1''',\bar{2}2'}&=& \sum_{1',1''}\gamma_{\bar{1}1',
   \bar{1}''1'''}G^{off}_{\bar{1}'2'} G^{off}_{\bar{2}1''}\, ,\\
 \label{eq:ee-mult} 
 \kappa^{ee}_{\bar{1}\bar{1}'',22'}&=& \sum_{1',1'''}\gamma_{\bar{1}1',
   \bar{1}''1'''}G^{off}_{\bar{1}'2'} G^{off}_{\bar{1}'''2}\, .
\end{eqnarray}
\end{mathletters}
The three functions $\kappa^\alpha$ form three integral kernels of the
Bethe-Salpeter equations for three vertex functions $\Gamma^\alpha$.  We
hence have three different nonlocal vertex functions being of the same
order in the small parameter $d^{-1}$. These functions differ in their
nonlocal parts and reduce to the same vertex $\gamma$ when all four end
points are from the same site. The full vertex function is then a sum of
the solutions to the three Bethe-Salpeter equations from which the local
vertex $\gamma$ must be twice subtracted.

It is more instructive to give an explicit representation for the full
two-particle vertex function in the Fourier representation with fermionic
$k=({\bf k},i\omega_n)$ and bosonic $q=({\bf q},i\nu_m)$ four-momenta. The
three contributions to the full vertex function in $k$-space are obtained
by Fourier transforming (\ref{eq:BS-lambda}). Two-particle bubbles
\begin{equation}
  \label{eq:X-def}
  \chi^{\pm}_{\sigma\sigma'}({\bf q};i\omega_n,i\omega_{n'})=\frac
  1N\sum_{\bf k} G_{\sigma}({\bf k},i\omega_n)G_{\sigma'}({\bf q}\pm{\bf
    k},i\omega_{n'}) 
\end{equation}
carry the surviving momentum dependence of the two-particle functions in
the infinite-dimensional limit. We have an explicit representation
\cite{Brandt89} for a hypercubic lattice with the nearest-neighbor
hopping $t=1/\sqrt{d}$
\begin{eqnarray}
  \label{eq:chi}
  &&\chi^{\pm}_{\sigma\sigma'}({\bf q};i\omega_n,i\omega_{n'})=
  -\mbox{sign}(\omega_n \omega_{n'})\!\int_{-\infty}^\infty
  \!\!\!d\lambda d\lambda' \theta(\lambda\omega_n) \theta(\lambda'
  \omega_{n'}) \nonumber \\ && \exp\left\{i\left(\lambda
    x_\sigma(i\omega_n)   +\lambda' x_{\sigma'}(i\omega_{n'})\right)
  -\frac 14\left(\lambda^2+\lambda'^2+2\lambda\lambda' X({\bf q})\right)
\right\}  \nonumber \\
\end{eqnarray}
with $x_\sigma(z)=z+\mu+\sigma B -\Sigma_\sigma(z)$
and $X({\bf q})=d^{-1}\sum_{\nu=1}^d \cos( q_\nu)$. The last quantity,
measuring nonlocality in infinite dimensions, behaves as a Gaussian random
variable with variance $d^{-1}$ when summed over momenta. The local bubble,
entering the local vertex $\gamma$, is a product of local propagators and
equals $\chi$ with $X=0$.

\begin{figure}
  \epsfig{figure=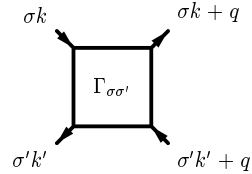,height=25mm}
\caption{\label{fig:2P-generic} Generic vertex function in $k$-space
  with three independent four-momenta and a defined order of incoming and
  outgoing fermions.}
\end{figure}
To clarify the meaning of different four-momenta in the vertex
functions we introduce a generic notation shown in
Fig.~\ref{fig:2P-generic}. The full vertex function in leading asymptotic
$d^{-1}$  order in the absence of external sources $J_\alpha$ reads in this 
notation
\begin{eqnarray}
  \label{eq:vertex-full}
 && \Gamma_{\sigma\sigma'}({\bf k},i\omega_n;{\bf k}',i\omega_{n'}; {\bf
    q},i\nu_m)=\Gamma_{\sigma\sigma'}^U({\bf q};i\omega_n,i\omega_{n'},
  i\nu_m)   \nonumber\\[2pt] 
 && +\Gamma_{\sigma\sigma'}^{ee}({\bf k}+{\bf k}'+{\bf q};i\omega_n,
 i\omega_{n'},i\nu_m) -\gamma_{\sigma\sigma'}(i\omega_n,i\omega_{n'},
  i\nu_m)   \nonumber \\[2pt]
&&+\Gamma_{\sigma\sigma'}^{eh}({\bf k}-{\bf  k}';i\omega_n,i\omega_{n'},
 i\nu_m)-\gamma_{\sigma\sigma'}(i\omega_n,i\omega_{n'},
  i\nu_m) 
\end{eqnarray}
where the each momentum-dependent contribution is determined from a
separate Bethe-Salpeter equation (\ref{eq:BS-lambda}) with the appropriate
multiplication from (\ref{eq:multiplication}).  Each of the nonlocal terms
in (\ref{eq:vertex-full}) depends only on one transfer momentum reflecting
the fact that maximally two different lattice sites are relevant. However,
because of different ways to connect the end points in the two-particle
vertex, the full function depends on all three independent momenta.  We
have $\gamma(\ldots)=N^{-1}\sum_{\bf q}\Gamma^\alpha({\bf q};\ldots)$ for
all channels $\alpha=U,eh,ee$.

To construct a Luttinger-Ward functional $\Phi$ generating higher-order
functions of the mean-field theory, we have to account properly for the
off-diagonal propagators.  Unlike the diagonal elements, suffering no
restriction in $d=\infty$, the off-diagonal propagators contribute only
under favorite circumstances, i.~e.  when properly combined with summations
over neighboring sites. That is why the off-diagonal elements appear in a
\textit{perturbative} manner in the mean-field limit. However, going to
higher-order Green functions we have to take into account higher powers of
the off-diagonal elements. The two-particle functions are generally of
order $d^{-1}$. They can be generated via functional derivatives of the
local, but inhomogeneous self-energy, or by derivatives of the nonlocal
self-energy w.r.t. nonlocal sources or the off-diagonal propagator. The
leading order of the nonlocal part of the self-energy is $d^{-3/2}$ and of
the off-diagonal propagator $d^{-1/2}$.  Hence, a functional derivative
$\delta\Sigma^{off}/\delta G^{off}\propto d^{-1}$ contributes in leading
order to the two-particle vertex. To turn the mean-field theory a
$\Phi$-derivable approximation, we have to consider \textit{nonlocal}
external sources $J^{off}$ connecting fluctuating fields at different
lattice sites.  The two-particle functions not derivable with local
external sources can then be generated via nonlocal derivatives
\begin{equation}
  \label{eq:2P-nonlocal}
   L^{eh}_{\bar{1}1',\bar{2}2'}=\eta\frac{\delta G_{\bar{1}2'}}{\delta
     J^{off}_{\bar{2}1'}}\, ,\,\, L^{ee}_{\bar{1}\bar{1}',22'}=\eta\frac{\delta
     G_{\bar{1}2}}{\delta J^{off}_{\bar{1}'2'}} \, .
\end{equation}
Generally, $n$-particle functions demand to keep nonlocal sources in
$\Phi$ up to the $n$th power. Once we introduce nonlocal external sources
into $\Phi$, we have to expand all quantities, including the one-particle
propagator and the self-energy appearing in $\Phi$ in powers in $J^{off}$. 

This perturbative construction with off-diagonal external sources is
applicable only in the high-temperature phase. At low temperatures, each
constituent of the two-particle vertex function (\ref{eq:vertex-full})
controls a different type of fluctuations that may cause phase transitions.
A singularity in the function $\Gamma^U$ signals a longitudinal, in
$\Gamma^{eh}$ a transversal (magnetic) order, and a singularity in
$\Gamma^{ee}$ indicates a superconducting instability.  Neglecting any of
the components of the full vertex function means suppressing potentially
relevant fluctuations.

The latter two transitions give rise to anomalous Green functions in the
ordered phase. To describe these phases correctly we have to introduce
anomalous (complex) local sources into the partition sum connecting, in a
hermitian manner, $\varphi_\uparrow^* \varphi_\downarrow^{\phantom{*}}$ for
the transversal order and $\varphi_\sigma^* \varphi_{\sigma'}^*$ in the
superconducting phase \cite{Janis99}. These anomalous perturbations induce
local anomalous propagators and self-energies as variational functions in
the generating functional \cite{note1}.  The mean-field Luttinger-Ward
functional, that is able to generate the complete leading asymptotics of
higher-order Green functions and to describe all low-temperature phases,
must contain either regular (of density type) nonlocal, or anomalous
(complex) but local external sources and their conjugate variational
functions.  However, only \textit{local} variational functions and order
parameters are treated nonperturbatively in $d=\infty$.

The newly derived vertex function (\ref{eq:vertex-full}) contains the full
momentum dependence and fulfills all the symmetry transformations of the
exact solution. It means that the current-current correlation function and
consequently the electrical conductivity \textit{do have} nontrivial vertex
corrections contrary to general expectations \cite{Khurana90,Georges96}. It
is only the function $\Gamma^{eh}$ that looses vertex corrections to the
conductivity due to the symmetry. The other two, explicitly depending on
${\bf q}$, correct the conductivity of a single particle-hole bubble in
high dimensions.

To conclude, we showed how to construct a generating functional for the
mean-field theory with the correct asymptotics of higher-order Green and
correlation functions. The mean-field theory as a complete solution in
$d=\infty$  is a $\Phi$-derivable theory if either regular nonlocal or
anomalous local sources are introduced.  Only then we are sure to generate
the complete leading asymptotics of higher-order vertex functions and to
describe all possible low-temperature phases.  However, the full
higher-order vertex functions in $d=\infty$ cannot be generated from single
equations or directly via functional derivatives from the Luttinger-Ward
functional as in the exact theory.

The work was supported by grant No. 202/98/1290 of the Grant Agency of the
Czech Republic.


\begin{references}
\bibitem{Baym61} G.~Baym and L. P. Kadanoff, \newblock Phys. Rev. {\bf
    124}, 287 (1961).
  
\bibitem{Baym62} G.~Baym, \newblock Phys. Rev. {\bf 127}, 1391 (1962).
  
\bibitem{Metzner89} W.~Metzner and D.~Vollhardt, \newblock Phys. Rev. Lett.
  {\bf 62}, 324 (1989).
  
\bibitem{Mueller-Hartmann89} E.~M\"uller-Hartmann, \newblock Z. Physik
  B{\bf 74}, 507 (1989).
  
\bibitem{Janis91} V.~Jani\v{s}, \newblock Z. Phys. B {\bf 83}, 227 (1991).
  
\bibitem{Janis92b} V.~Jani\v{s} and D.~Vollhardt, \newblock Phys. Rev.
  B{\bf 46}, 15172 (1992).
  
\bibitem{Georges96} A.~Georges, G.~Kotliar, W.~Krauth, and M.~Rozenberg,
  \newblock Rev. Mod. Phys. {\bf 68}, 13 (1996).
  
\bibitem{Brandt89} U. Brandt and C. Mielsch, \newblock Z. Physik B{\bf 75},
  365 (1989)
  
\bibitem{Engelbrecht95} J.~R.~Engelbrecht and K.~S.~Bedell, \newblock Phys.
  Rev. Lett. {\bf 74}, 4265 (1995) and K.~S.~Bedell, J.~R.~Engelbrecht, and
  K.~B.~Blagoev, \newblock preprint \textit{cond-mat}/9808203.
  
\bibitem{vanDongen98} P.~G.~J.~van~Dongen, G.~S.~Uhrig, and
  E.~M\"uller-Hartmann, \newblock preprint \textit{cond-mat}/9807276.
  
\bibitem{note0} Generally, maximally $n$ distant lattice sites are relevant
  for $n$-particle quantities.
  
\bibitem{Janis99} V.~Jani\v{s}, submitted to Phys. Rev. B.
  
\bibitem{note1} The anomalous (local) Green functions vanish at equilibrium
  in the high-temperature phase and serve as order parameters at
  low-temperatures. They can also alternatively be used to generate the
  higher-order vertex functions.
  
\bibitem{Khurana90} A.~Khurana, \newblock Phys. Rev. Lett. {\bf 64}, 1990
  (1990).

\end{references}
\end{document}